\documentstyle[epsfig]{mn}

\title[Photo-chemical evolution of elliptical galaxies]
{Photo-chemical evolution of elliptical galaxies II. The impact of merging-induced starbursts}

\author[A. Pipino \& F. Matteucci]{Antonio Pipino and Francesca Matteucci
\\
Dipartimento di Astronomia, Universit\`a di Trieste,
    Via G.B. Tiepolo 11, I-34127, Trieste, Italy}

\date{Accepted,
      Received }

\begin{document}
\maketitle

\begin{abstract}
The effects  
of late gas accretion episodes and subsequent merger-induced starbursts
on the photo-chemical evolution of elliptical galaxies are studied and compared to the picture of galaxy formation
occurring at high redshift with an unique and intense starburst modulated by a very short infall, 
as suggested by Pipino \& Matteucci (2004, Paper I).
By means of the comparison with the the colour-magnitude relations
and the $[<Mg/Fe>_V]$-$\sigma$ relation observed in ellipticals, we conclude that either bursts involving
a gas mass comparable to the mass already transformed into stars during the first episode of star formation
and occurring at any redshift, or bursts occurring at low redshift (i.e. z$\le$0.2) and with a large
range of accreted mass,
are ruled out.
These models fail in matching the above relations even if 
the initial infall hypothesis is relaxed, and the galaxies form either by means of more
complicated star formation histories or  
by means of the classical monolithic model. On the other hand,
galaxies accreting a small amount of gas at high redshift (i.e. z$\ge$3) produce a spread in the model results,
with respect to Paper I best model, which is consistent with the observational scatter
of the color-magnitude relations, although there is only marginal agreement with the $[<Mg/Fe>_V]$-$\sigma$ relation.
Therefore, only small \emph{perturbations} to the standard scenario
seem to be allowed.
We stress that the strongest constraints to galaxy formation mechanisms are 
represented by the chemical abundances, whereas the colours can be reproduced under
several different hypotheses.

\end{abstract}


\section{Introduction}

The debate on the formation of spheroids is still one of the major issues in astrophysics.
In particular, the monolithic scenario (e.g. Larson, 1974, Arimoto \& Yoshii, 1987, Matteucci \& Tornambe', 1987)
and the hierarchical paradigm (e.g. White \& Rees, 1978, Kauffmann \& Charlot 1998, Steinmetz \& Navarro 2002, 
Meza et al. 2003) are the main competing theories 
\emph{on the market}. 

According to the monolithic scheme,
ellipticals are assumed to have formed at high redshift as a result of a rapid collapse of a gas cloud,
characterized by a short and intense starburst, followed by a galactic wind at very early epochs.
This scenario naturally explains the observations of nearby ellipticals showing a
Mass-Metallicity relation (e.g. Carollo et al.,
1993; Gonzalez, 1993; Davies et al., 1993; Trager et al. 1998, 2000)
and a reddening of the stellar light (Color-Magnitude relation, hereafter CMR,
Bower et al., 1992) with the velocity dispersion of the galaxies, because the most massive ellipticals, having
the deepest potential wells, are the most metal rich. 
The lack of significant change in the slope and in the scatter of the CMR (Stanford et al., 1998, but see Kaviraj et al., 2004),
the slow evolution of colors (e.g. Saglia et al. 2000; Ellis et al., 1997), mass to light
ratios (e.g. van Dokkum $\&$ Franx, 1996; van Dokkum \& Ellis 2003; Rusin et al., 2003; van de Ven et al., 2003) and line strength indices
of cluster early-type galaxies out to $z\sim 1$ (e.g. Bernardi et al., 2003; van Dokkum et al. 1998),
are consistent with a passive aging of the stellar populations and hint toward a high and quite synchronized
redshift of formation for these galaxies. 
Further support to this picture seems to emerge from the very recent observations
which lead Cimatti et al. (2004) to the identification of four old, fully assembled, $10^{11}M_{\odot}$ spheroids at 1.6 $< z <$ 1.9.

On the other hand, hierarchical semi-analytic models predict that
ellipticals are formed by mergers of already formed stellar systems,
which closely follow the behaviour of their dark matter haloes 
and regulate the chemical enrichment of the system (White $\&$ Rees, 1978). In this picture
massive ellipticals form at relatively recent epochs through major mergers between
spiral galaxies (e.g. Kauffmann $\&$ White, 1993) occurring over a large redshift interval. 
Evidences favouring the hierarchical scenario are represented by the observed interacting 
galaxies (on-going mergers), ellipticals with disturbed morphologies
(i.e. counter-rotating cores, dust lanes, ripples, e.g. Kormendy $\&$ Djorgovski, 1989),
the morphology-density relation in clusters (Dressler et al., 1997), the relatively large values of
the $H_{\beta}$ index in a non negligible fraction of nearby ellipticals (Gonzalez 1993, Trager et al 1998).

We refer the reader to Paper I, Peebles (2002), Calura et al. (2004)
for a more detailed discussion on the topic and references. Here we recall that
in Paper I we showed that the majority of the optical properties of elliptical galaxies
can be simultaneously reproduced under the assumption that
the formation process, occurring at high redshift (as in the monolithic collapse scenario), 
is stronger and faster in more massive objects
with respect to the less massive ones.
In fact, the $[<Mg/Fe>_V]$ ratio in the cores of ellipticals increases with galactic mass 
(Worthey et al. 1992; Weiss et al. 1995; Kuntschner 2000, Kuntschner et al., 2001), being always larger than zero
in the core of bright galaxies (e.g. Faber et al., 1992; Carollo et
al., 1993; Davies et al., 1993; Worthey et al., 1992). This strongly suggests 
that the star formation (SF) lasted for a period shorter than the time at which the Fe pollution from type 
Ia SNe becomes dominant (see Weiss et al., 1995, Matteucci, 1994). 
Recently, a similar suggestion was proposed by Granato et al. (2004)
in the scheme of the joint formation of ellipticals and QSOs.

In order to complete the study, we aim at verifing whether Paper I conclusions
can be achieved with a more complicated formation history, namely to study in a simple way the effect
of merger induced starbursts on the properties of a dominant stellar population assembled 
at high redshift.
A complementary view on the analysis made in this paper is presented by Thomas (1999) and Thomas et al. (1999).
In particular, they compared the outcome of a fast collapse
to the results obtained by the merger of two spirals and found difficult to reproduce the $[<Mg/Fe>_V]$
overabundance in the latter case, unless very flat IMFs are invoked during the merger. Furthermore,
since the $[<Mg/Fe>_V]$-$\sigma$ relation seems to be independent from the environment and correlates
with galactic age (the older is the galaxy the higher is its $[<Mg/Fe>_V]$), 
Thomas et al (2002) concluded that SF histories (SFHs) based on the hierarchical clustering
scenario (Kauffmann \& Charlot, 1998) cannot reproduce the observed chemical properties of elliptical galaxies.
On the other hand, by means of broad-band photometry alone,
Shioya \& Bekki (1998) constrained the epoch of the major merger. They found that mergers between 
spirals are allowed even at z=0.3, but their models cannot account for
the Mg enhancement.
Therefore, it seems necessary to couple the information coming from the chemistry
to those from the broad-band colours in order to better constrain the SFH of spheroids.
Due to the fact that we are not modelling the dynamical evolution, our treatment of the galaxy merging is quite simple
and focused only to study how the mean chemical and photometric properties of the stellar populations
change as a consequence of late episodes of star formation triggered by gaseous mergers.
In particular, we analysed different scenarios involving high mass spheroids (i.e.  initial gas mass \rm 
$M_{lum}=10^{11-12}M_{\odot}$),
as predicted by Paper I's best model, that we consider
our \emph{fiducial case}, and we took it as a basis for studying 
simulated accretion of primordial gas triggering a starburst.
We explored several cases by varying the duration of the starburst and its intensity.

This paper is organized as follows: in section 2 we present the model. 
In section 3 we discuss the results and in section 4 we draw some conclusions.
In order to be consistent with the results of Paper I, we assume 
a cosmological model with $\Omega_m=0.3$, $\Omega_{\Lambda}=0.7$
and $H_o=70 \rm \,km\,  s^{-1}\, Mpc^{-1}$. For all models we
assume that the star formation starts at redshift $z_f$=5. Variations
in $z_f$ do not significantly affect the conclusions.

\section{The model}

The adopted chemical evolution model is based on that
presented in Paper I. In this particular case, however, we consider our model galaxies
as a single zone extending out to one effective radius, 
with instantaneous mixing of gas. Moreover
we take explicitly into account a possible mass flow due to the galactic wind
and a possible secondary episode of gas accretion. Therefore, the
equations of chemical evolution take the following form:
\begin{eqnarray}
{d G_i (t) \over d t}  =  -\psi (t) X_i (t)\,   \nonumber \\
 +\int_{M_L}^{M_{B_m}} \psi (t-\tau_m) Q_{\rm mi}(t-\tau_m) \phi (m) dm\,  \nonumber \\
 + A\int_{M_{B_m}}^{M_{B_M}} \phi (m) \left [ \int_{\rm\mu_{\rm min}}^{0.5} f(\mu) Q_{\rm mi}(t-\tau_{\rm m_2}) \psi (t-\tau_{\rm m_2}) d\mu \right ] dm \,  \nonumber \\
  +(1-A)\, \int_{\rm M_{\rm B_m}}^{M_{\rm B_M}} \psi (t-\tau_m) Q_{\rm mi}(t-\tau_m) \phi (m) dm\,  \nonumber \\
  +\int_{\rm M_{\rm B_M}}^{M_U} \psi (t-\tau_m) Q_{\rm mi}(t-\tau_m) \phi (m) dm\,  \nonumber \\
 +({d G_i (t) \over d t})_{\rm infall}-W(t)X_i (t)
+({d G_i (t) \over d t})_{\rm acc}\,  ,	
\label{dgdt-eq}
\end{eqnarray}
where $G_i (t)$ is the normalized mass density
of the element \emph{i} at the time \emph{t} in the ISM.
$X_i (t)$ 
is defined as the abundance by mass of the element \emph{i}.
We address the reader
to the original papers of Matteucci $\&$ Greggio (1986),
Matteucci $\&$ Tornambe' (1987),
Gibson (1997) and Paper I 
for a comprehensive discussion of these equations. 
By means of this equation we can calculate the evolution of 21 elemental species.
In particular, the integrals at the right-hand side of the equations give the 
rate at which the element \emph{i} is restored into the interstellar medium
both as newly-synthesized and already existing element by low- and intermediate-mass 
stars, Type Ia and Type II supernovae (SNe Ia and SNe II, respectively).
Here we recall that for SNe Ia we assume as a progenitor model 
the single degenerate one: a C-O white dwarf which accretes material from a companion
(the secondary), eventually reaching the Chandrasekar 
mass and exploding via carbon-deflagration (Whelan $\&$ Iben 1973). 
We make use of the SNe Ia rate formalism derived by
Greggio $\&$ Renzini  (1983).
The predicted SNe Ia explosion rate is constrained
to reproduce the present day observed value in ellipticals (Cappellaro
et al.\ 1999). This can be done by means of the parameter $A$ in eq. (1), which 
represents the fraction of binary systems in the IMF able to give rise to SNIa.

\subsection{The initial burst of star formation}

We refer to Paper I Model II input parameters for our \emph{fiducial model}, 
namely Salpeter (1955) IMF, Thielemann et al (1996) yields for massive stars,
Nomoto et al (1997) yields for type Ia SNe and van den Hoek \& Groenewegen (1997) yields for low-
and intermediate-mass stars.
According to this model, the infall timescale ($\tau$) and the star formation efficiency ($\nu$)
decreases and increases with increasing galactic mass, respectively.

\subsection{The merger-induced starburst}

In this paper we need to take into account the possible presence of a second infall
episode (which simulates the merger), 
therefore we added a \emph{merging} term in the right-hand side of the chemical evolution equation:
\begin{equation}
({d G_i (t) \over d t})_{acc}= X_{i,acc} C e^{-{t-t_{acc} \over \tau_{acc}}}\, .
\end{equation}
$X_{i,acc}$ describes the chemical composition of the accreted gas, assumed to be primordial.
This is a good assumption for all practical purposes, since, even in the case of gas with
average Intergalactic Medium metallicity ($Z\simeq 0.04 Z_{\odot}$, e.g. Calura and Matteucci, 2004), 
the metal abundances of the accreted 
gas are much lower than the ones in the ISM.  
C represents the normalization constant, evaluated by integrating the law from $t_{acc}$, namely the time
at which the merging episode starts, to the present time, and requiring that the galaxy accretes $M_{acc}$ solar masses
of gas.
Finally, $\tau_{acc}$ is the gas accretion timescale, fixed to be 0.2 Gyr. The reason for this choice
is linked to the fact that we expect a $\tau_{acc}\sim\tau$ for accreting a gas mass $M_{acc}\sim M_{lum}$.
In fact, we recall that $\tau\sim0.2-0.4$ Gyr in Paper I's best model.
Small variations in $\tau_{acc}$, however, do not affect our results.
Only a much longer timescale ($\tau_{acc}\sim 1Gyr$) would imply that  most of the newly accreted gas cannot be 
consumed during the second SF
episode.\rm

We recall that, once the conditions for the galactic wind onset are satisfied in a given time-step,
we allow the galaxies to lose all the residual gas present at that time. Only during the merging-induced
starburst, the mass flow rate $W(t)$ entering eq. (1) is assumed to be proportional to the star formation rate ($W(t)=\psi(t)$)
as suggested for starbursting galaxies (e.g. Heckman, 2002). It can
be shown that these assumptions do not dramatically alter our conclusions. A more detailed
treatment for the energetics and the mass exchange after the galactic wind has been developed in a separate 
paper (Pipino et al. 2005).

We limit our analysis to a one-zone treatment out to one effective radius, 
in such a way that the accreted gas is uniformly distributed in this region.
In this paper, we are interested to the effect of star formation histories
made by multiple episodes,
therefore we force the starburst to happen at the beginning of the merger (i.e. at time $t_{acc}$).
We stress that this condition is likely to hold, since we are dealing
with a large amount of infalling gas, and thus, strong cooling.

Since we are dealing not only with old stellar populations, the light originated by the stellar
populations born during the second burst could be a significant fraction of the total flux coming from the galaxy,
thus influencing the spectral features. Therefore we calculate the mean stellar abundances by weighting on the V luminosity. In particular, we 
follow Arimoto \& Yoshii (1987) and Gibson (1997), by means of the photometric code
by Jimenez et al. (1998). In any case, we confirm the previous findings (Matteucci et al. 1998; Gibson 1997; Thomas et al 1999)
that for old and massive objects the mass weighted averages are very close to the luminosity weighted ones, whereas, in the case
of more recent star formation, the former are systematically $\sim 0.2$ dex higher than the latter.

\rm
Once we have defined the parameters governing the merging
event, it is useful to separate the models analysed
in different groups, each one representing a different region in the
space defined by the parameters $\tau\, ,\tau_{acc}\, ,M_{lum}\, ,M_{acc}\, ,\nu$.
A simple classification can be done by considering the amount of stars formed
during the SF episodes.

\begin{itemize}
\item[Case 1:] Models in which the SFH of the main episode creates many more stars than the SF during the merger-induced burst.

\item[Case 2:] Models in which the SFH of the main episode is comparable to the SFH
during the merger-induced burst.

\item[Case 3:] Models in which the SFH of the main episode is less efficient than the SFH
during the merger-induced burst.
\end{itemize}

We will start from Case 1, showing that those models are "perturbations"
of the Paper I best model, and in many cases predict properties still in agreement
with the observations (Sec. 3.1). 
Other models belonging either to Case 1 or 2 will be presented in
sections 3.2 - 3.4. In particular, we 
verify whether a best model with merger-induced starburst 
which reproduces the properties of ellipticals can be found (Sec. 3.2 and 3.3).

\rm

\section{Results and discussion}

\subsection{Perturbing the best model}

\begin{table*}
\centering

\begin{minipage}{120mm}
\begin{flushleft}
\tiny
\caption[]{Merger parameters and model predictions}
\begin{tabular}{llclccclllll}
\hline
\hline

Model &$M_{lum}$&${M_{acc}\over M_{lum}}$& $t_{acc}$&$[<{Mg\over Fe}>_V]$&  $[<{Fe\over H}>_V]$&$[<{Mg\over H}>_V]$ &$M_V$&	U-V&	V-K&	J-K& $R_*$\\
&($M_{\odot}$) & &(Gyr) &  \\
\hline
Fiducial\\
\hline
a (Paper I)&$10^{11}$ & 0  &	0   &0.448	 &-0.284	 &0.049	 &-19.2	 &1.39	 &3.10	 &1.03	& 0	\\					

b&$10^{11}$ & 1  &	1   &0.292	&0.288	&0.153	&-19.8	&1.45	&3.24	&1.07&0.94	\\
c&$10^{11}$ & 1  &	10   &0.133	 &0.946	 &0.455	 &-20.2	 &1.13	&2.66	&0.91&0.54	\\
d&$10^{11}$ & 0.1  &1   &0.391	&0.553	&0.284	&-19.4	&1.48	&3.24	&1.07&0.33	\\
e&$10^{11}$ & 0.1  &10   &0.389	&0.330	&0.126	&-19.4	&1.32	&3.02	&1.01&0.11	\\

f (Paper I)&$10^{12}$ & 0  &	0   &0.433	&-0.129	&0.124	&-22.0	&1.48	&3.28	&1.09 & 0\\

g&$10^{12}$ & 1  &	1   &0.289	&0.372	&0.235	&-22.6	&1.57	&3.42	&1.13 & 0.26\\
h&$10^{12}$ & 1  &	10   &0.171	&0.843	&0.394	&-23.0	&1.23	&2.89	&0.98 & 0.74\\
i&$10^{12}$ & 0.1  &1   &0.415	&0.129	&0.145	&-22.0	&1.49	&3.30	&1.09 & 0.07\\
j&$10^{12}$ & 0.1  &10   &0.380	&0.330	&0.176	&-22.2	&1.41	&3.21	&1.07&0.11\\
\hline
Closed Box\\
\hline
k (CB)&$10^{11}$ & 0  &	0   &0.761	&-0.896	&-0.450	&-20.0&	0.76	&2.73&	0.97 & 0\\

l&$10^{11}$ & 1  &	1   &0.524	 &0.168	 &0.089	 &-20.0	 &1.28	 &3.14	 &1.04 &0.46\\
m&$10^{11}$ & 1  &	10   &0.355	 &0.840	 &0.335	 &-20.5 &1.09	 &2.69 &	0.91 &1.00\\
n&$10^{11}$ & 0.1  &1   &0.735	 &-0.196	 &-0.064	 &-19.7 &	1.24	 &3.11	 &1.04 &0.07\\
o&$10^{11}$ & 0.1  &10   &0.674	 &0.349	 &0.031	 &-19.7	 &1.12	 &2.90	 &0.97 &0.08\\

p (CB)&$10^{12}$ & 0  &	0   &0.900	 &-0.167	 &-0.015	 &-22.1	 &1.11	 &2.97	 &1.00 & 0\\

q&$10^{12}$ & 1  &	1   &0.740	 &0.388	 &0.143	 &-22.5	 &1.30	&3.26	&1.08&0.98\\
r&$10^{12}$ & 1  &	10   &0.573	 &0.844	 &0.279	 &-23.0	 &1.09	 &2.77	 &0.94&0.97\\
s&$10^{12}$ & 0.1  &1   &0.889	&0.080	&0.001	&-22.1	&1.13	&3.01	&1.01&0.09\\
t&$10^{12}$ & 0.1  &10   &0.847	&0.344	&0.036	&-22.2	&1.11	&2.94	&0.99&0.14\\
\hline
Miscellanea\\
\hline
a+AY87&$10^{11}$ & 0  &	0   &0.614	&0.269	&0.820	&-19.8	&2.15	&3.85	&1.24  & 0	\\
b+AY87&$10^{11}$ & 1  &	1   &0.431	&0.855	&1.070	&-20.4	&1.41	&3.02	&1.04&1.04	\\
c+$\nu$=20&$10^{11}$ & 1  &	10   &0.274	&0.090	&0.221	&-20.9	&1.01	&2.69	&0.91&1.85	\\
b+$t_{burst}$=0.5&$10^{11}$ & 1  &	1   & 0.305& 0.081&0.109&-19.8&1.43&3.21&1.06&0.65\\
b+$t_{burst}$=0.1&$10^{11}$ & 1  &	1   & 0.417&-0.263&0.028&-19.3&1.36&3.06&1.01&0.01\\
c+$t_{burst}$=0.5&$10^{11}$ & 1  &	10   &0.293	&0.005	&0.194	&-20.4	&0.99	&2.63	&0.89&0.75	\\
c+$t_{burst}$=0.3&$10^{11}$ & 1  &	10   & 0.355&-0.137&0.088&-19.8&1.14&2.85&0.96&0.15\\
d+$t_{burst}$=0.5&$10^{11}$ & 0.1  &	1   &0.428 &-0.026 &0.098 &-19.3&1.34&3.05&1.02&0.01\\
e+$t_{burst}$=0.5&$10^{11}$ & 0.1  &	10   & 0.426&-0.216&0.073&-19.3&1.34&3.05&1.02&0.01\\
b+A=0.06&$10^{11}$ & 1  &	1   &0.372	&-0.012	&0.194	&-20.0	&1.51	&3.36	&1.11&0.35	\\
u($\nu$=55) &$10^{12}$ & 1  &	1   &0.363&	0.696&	0.388&	-22.2&	1.67 &	3.48 &	1.14& 0.48\\
v($\nu$=110) &$10^{12}$ &1  &	2   &0.416&	0.839&	0.416&	-21.8&	1.48 &	3.25 &	1.08&1.67\\
\hline
\hline
\end{tabular}
\end{flushleft}
Paper I's best model and closed box model
without mergers are labelled as \emph{Paper I} and \emph{CB}, respectively. 
\end{minipage}
\label{tab-mod}
\end{table*}

\begin{figure}
\epsfig{file=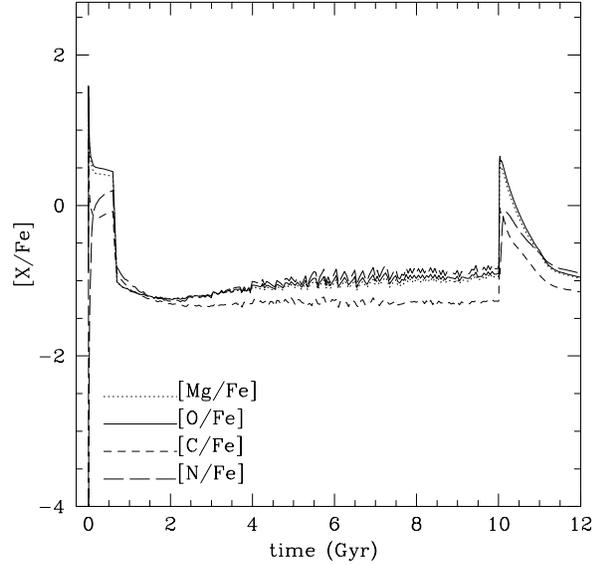, height=8cm, width=8cm}
\caption{Time evolution of abundance ratios in the ISM of a model \emph{c} galaxy. }
\label{fig-maio}
\end{figure}

\begin{figure}
\epsfig{file=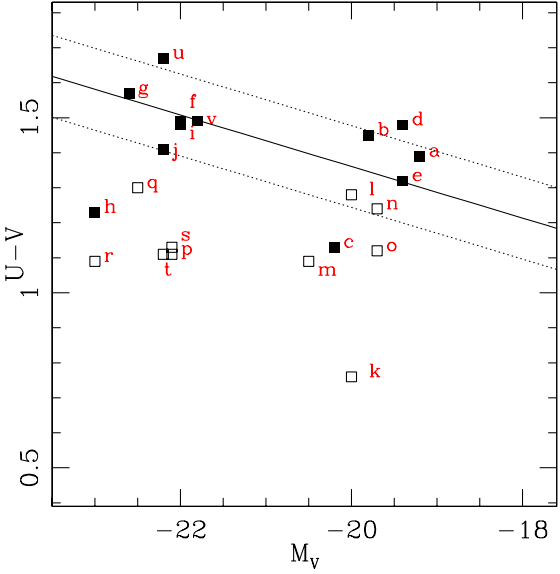, height=8cm, width=8cm}
\caption{Fiducial (full squares) and closed box (empty squares) model predictions
compared with observations. The mean value of the observed colour at a given $M_V$ is given by the
solid line (fit to Bower et al.'s, 1992, data). The observed scatter is
represented by the dashed lines (the offset from the mean value
is taken to be $\pm 3\sigma$, Bower et al., 1992).
Models are labelled according to Table 1. 
}
\label{fig-cmr-a}
\end{figure}

\begin{figure}
\epsfig{file=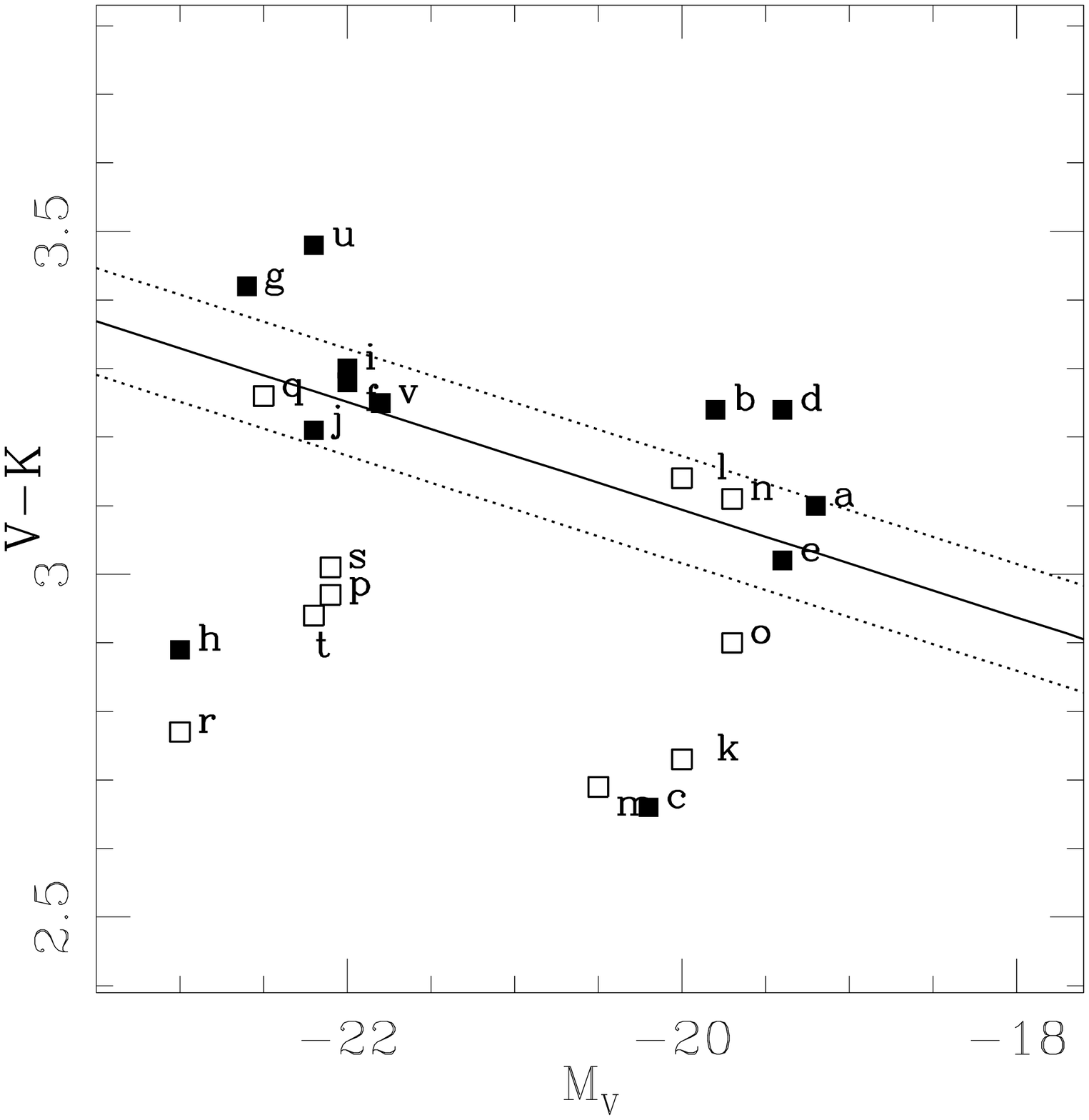, height=8cm, width=8cm}
\caption{Fiducial (full squares) and closed box (empty squares) model predictions
compared with observations. Fit to the data (solid) and 3$\sigma$ boundary (dashed) by Bower et al (1992).
Models are labelled according to Table 1.
}
\label{fig-cmr-b}
\end{figure}

\begin{figure}
\epsfig{file=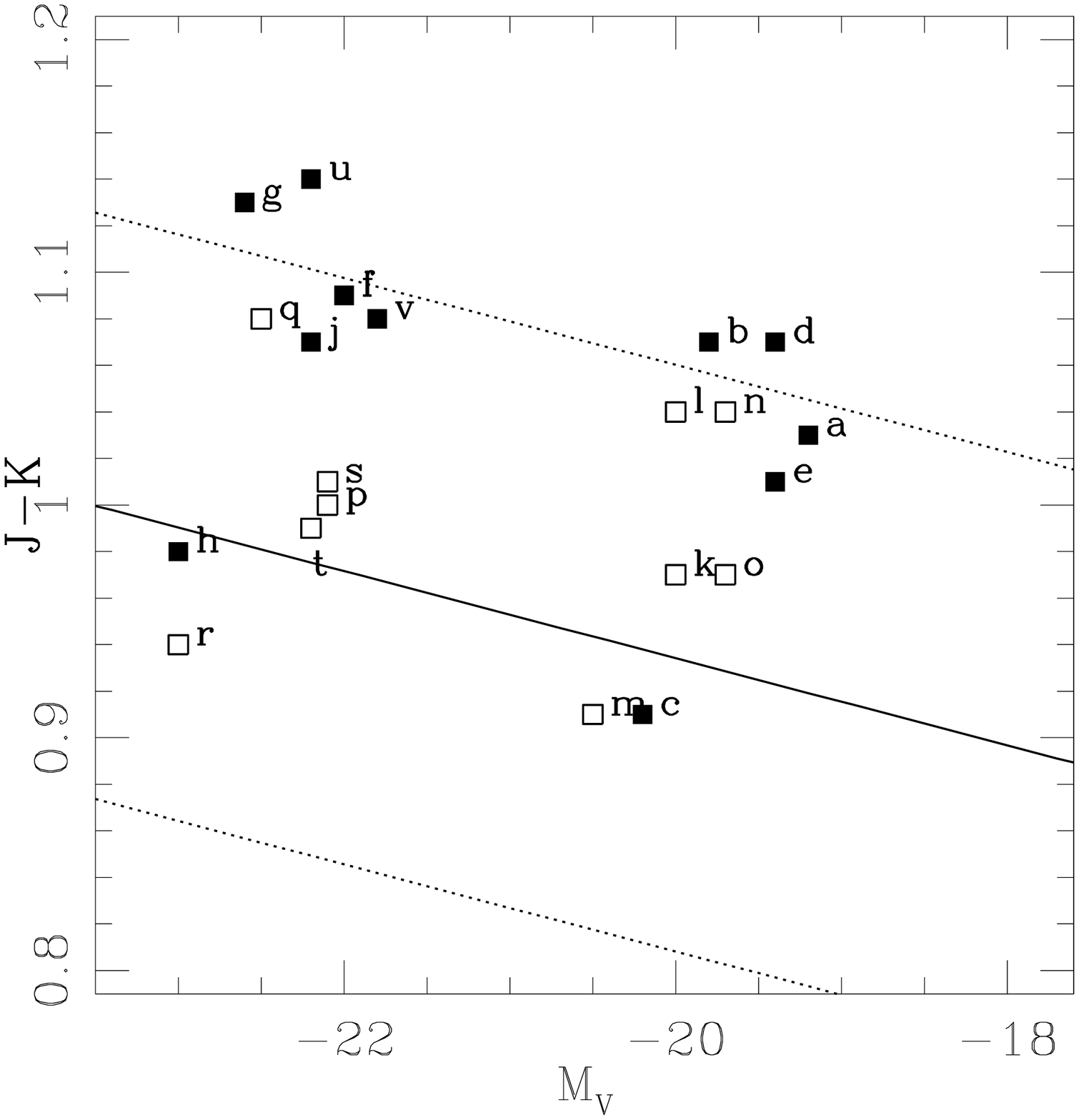, height=8cm, width=8cm}
\caption{Fiducial (full squares) and closed box (empty squares) model predictions
compared with observations. Fit to the data (solid) and 3$\sigma$ boundary (dashed) by Bower et al (1992).
Models are labelled according to Table 1. 
}
\label{fig-cmr-c}
\end{figure}

\begin{figure}
\epsfig{file=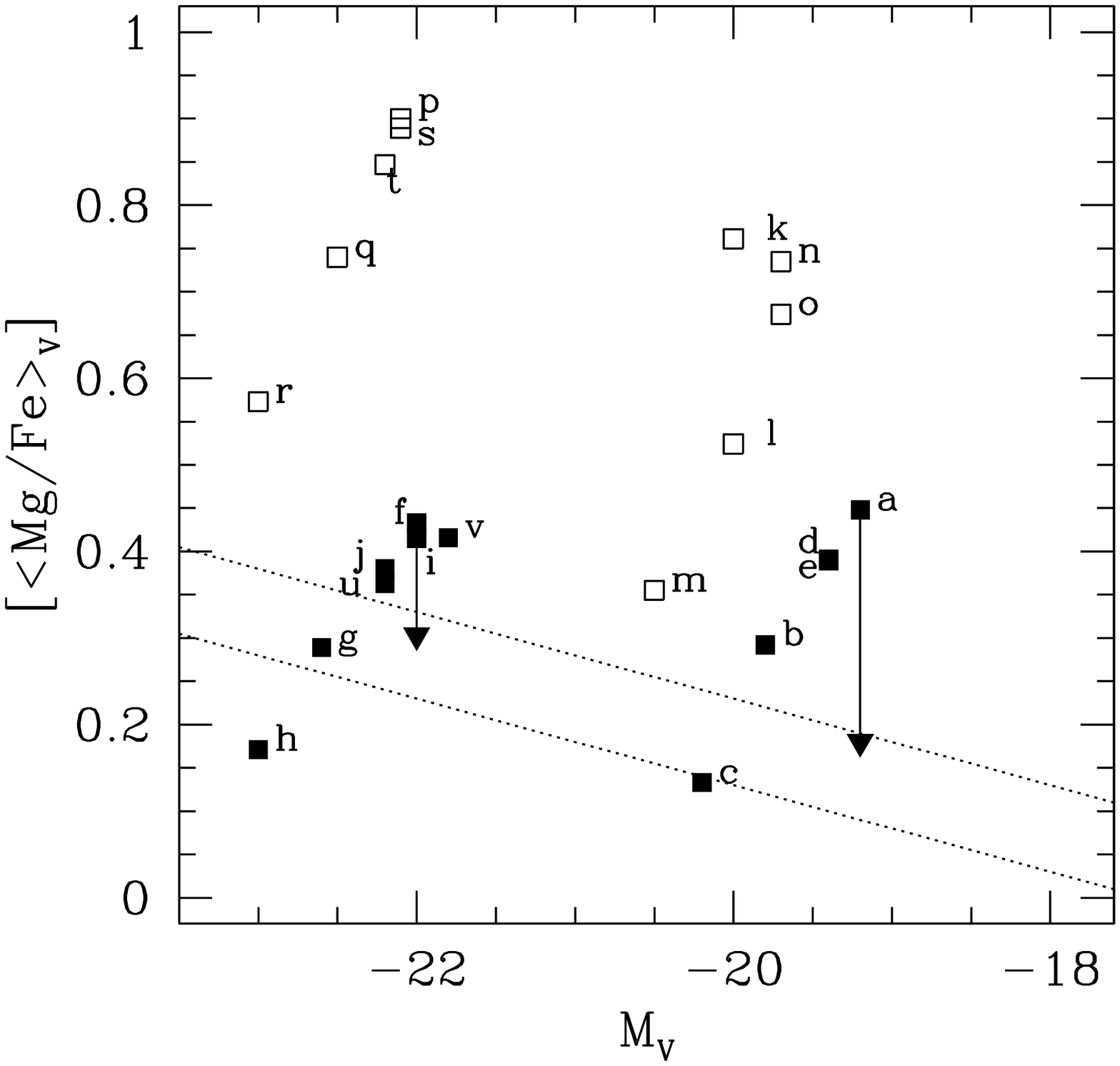, height=8cm, width=8cm}
\caption{Fiducial (full squares) and closed box (empty squares) model predictions
(entire galaxy) compared with observations (within $0.1 \rm R_{eff}$). The dashed
lines limit the observed values as given by Thomas et al (2002) from data 
by Gonzalez (1993), Mehlert et al. (2000), Beuing et al. (2002).
Models are labelled according to Table 1.
The arrows show that the predictions for the whole galaxies become consistent with observations,
when the radial gradient in $[<Mg/Fe>_V]$ (as predicted by Paper I's best model) is taken into account.
}
\label{fig-cmr-d}
\end{figure}

\rm
We start to explore the parameter space by considering
merger-induced starburst as perturbations of Paper I's best model.
\rm
In this case the merging-induced starburst is forced to last for $t_{burst}=$ 1 Gyr with the same star formation efficiency of the initial one 
(i.e. $\nu=10\rm Gyr^{-1}$ and  $\nu=22\rm Gyr^{-1}$ for  models with initial mass in gas \rm
$M_{lum}=10^{11}M_{\odot}$ and $M_{lum}=10^{12}M_{\odot}$, respectively),
unless otherwise stated. As a first step, in fact, the burst duration and efficiency in converting gas into stars
are chosen in order to have duration and intensity similar to those characterizing the main star formation episode (i.e. before the galactic wind). In this case, the SFH during the second burst will be influenced only by the \rm gas supply ($M_{acc}$).
 
We explored very extreme cases, namely we run models in which the  accreted gas mass is either $\sim 10\%$ (i.e.
those probably leading to Case 1) or $\sim 100\%$ (which, according to their SFH, will
belong either to Case 1 or 2) \rm
of the initial mass, occurring either 1 or 10 Gyr after the galactic evolution has started.
Mergers occurring at lower redshifts are not allowed, to avoid violation of  the constraints
on the present-day Type II SNe rate in ellipticals; in fact, SNII are not 
observed in local ellipticals (Cappellaro et al. 1999).

We present our results for the fiducial case in Table 1 (models \emph{a-j}), where, each galactic model is identified
by a letter and the initial luminous  gas \rm mass in Cols. 1 and 2, respectively. 
We show in Cols. 3 and 4 the amount of  gas \rm mass accreted during the merger and the time at which
the accretion begins, respectively; in Cols. 5-7 we present the
stellar $[<Mg/Fe>_V]$, $[<Fe/H>_V]$ and $[<Mg/H>_V]$. Finally, in Cols. 8-11 we show the V magnitude and the U-V,
V-K, J-K colours.

In the last column we present the ratio $R_*$ between the mass in stars 
assembled during the merger-induced burst and still alive at the present time, 
and the stars formed during the initial
SF episode. We consider the models in which $0.5 \le R\le 1.5$ belonging to Case 2. 
Before analysing the results, it is important to keep in mind that
models accreting the same amount of gas ($M_{acc}$) can undergo different SFHs. This is due to the fact that
when $t_{acc}$ is low the galaxy could be still undergoing a galactic wind. Therefore,
the net amount of gas available for new SF (i.e. the sum of the gas left after the initial burst, the gas 
restored by stellar mass loss and the newly accreted gas minus mass lost
for the wind) might be smaller than 
the ISM mass of a galaxy which, after the wind phase, passively evolves out to, e.g., $t_{acc}\sim 10$ Gyr.
In this last case, it is more likely the all the accretd mass $M_{acc}$ can be turned into stars.
\rm
The second group of rows (models \emph{k-t}) will be discussed in Sec. 3.2. \rm 
Other model belonging either to Case 2 or Case 3 are grouped under \emph{miscellanea} and
will be discussed in sections 3.3 and 3.4. \rm	
As it can be seen from the entries of this Table, the most evident effect produced by the merging-induced
starburst to the fiducial model is
a decrease in $[<Mg/Fe>_V]$, although the ISM abundances increase as a consequence of the increased stellar
nucleosynthesis. The reason lies in the fact that the [Mg/Fe] ratio in the ISM
after the galactic wind (and before the occurrence of the merger) is dominated by Type Ia SN explosions, which
keep its value well below the solar ratio. When the merger-induced starburst occurs, most stars form out of 
this gas (thus lowering the total $[<Mg/Fe>_V]$); at the same time, the new generations of Type II SNe
produce a sudden increase in the ISM [$\alpha$/Fe] ratios, but the value reached in this way is lower than the typical
abundance ratio of the ISM during the first $\sim$ 500 Myr (see Fig. \ref{fig-maio}). Therefore, the new SNII contribution is not
enough to make $[<Mg/Fe>_V]$ rise again significantly. 
Since the star formation efficiency and the infall time scale for the fiducial case were tuned to reproduce
the color-magnitude, the mass-metallicity and the $[<Mg/Fe>_V]$-$\sigma$ relations (see Paper I), the straightforward
result is that, at this stage, the merging episode worsens the results obtained in Paper I. 
We recall that the star formation rate peak values during either the first or the second burst have
similar intensity (e.g. $\sim 80M_{\odot}yr^{-1}$ in the case of models \emph{d}) .

\subsection{Comparison with merging-induced bursts in a Closed Box case}

A possible alternative route to reconcile the presence of mergers with
the observed $[<Mg/Fe>_V]$, requires perhaps more complex star formation histories.
As a first step,
we run also a \emph{closed box} model (Table 1, models \emph{k-t}), with the same input parameters of the fiducial case, 
the only difference being that
in this case the gas mass is already present since the beginning. From the analysis made in Paper I, in fact, we know
that this case gives very high $[<Mg/Fe>_V]$, at odds with observations. Therefore we are interested in
checking if a closed box model plus a second starburst triggered by a merger (which always lowers $[<Mg/Fe>_V]$ in the stellar populations),
can match the observations. In this way, the fiducial and the closed box model allow us to restrict a first 
region of the parameter space
\rm ($\tau$ varying in the interval [0,0.4] Gyr) \rm in which also intermediate cases can occur, possibly producing \emph{normal} ellipticals. 

In order to derive more quantitative constraints we compare both fiducial and closed box cases with observations
in Figs. \ref{fig-cmr-a}-\ref{fig-cmr-d}. In particular, the fiducial models (either a $10^{11}M_{\odot}$ or a $10^{12}M_{\odot}$ galaxy)
are represented by filled squares, whereas the closed box galaxies by empty ones. 
The solid (dashed) lines give the fit (3 $\sigma$ boundary) to the observed values in the case of the three CMRs, as
given by Bower et al (1992). In the case of the $[<Mg/Fe>_V]$-$M_V$ relation (Fig. \ref{fig-cmr-d}), the lines bracket the observed region
as given by Thomas et al (2002).

The fiducial model still reproduces the CMRs in most of the cases when mergers occur; the same happens in 
the semi-analytical models applied to the hierarchical clustering scenario, which
succeed in reproducing the CMRs (e.g. Kauffmann \& Charlot, 1998), but fail in reproducing the 
$[<Mg/Fe>_V]$-$\sigma$ relation, as shown by Thomas \& Kauffmann (1999).
Care should be taken since our predictions
are made for a one-zone model, thus neglecting radial gradients in the colours and in the abundances. Furthermore,
the observations were made by means of a fixed aperture which is in general different from the radius considered here (Bower et al, 1992;
see also Paper I). In any case, the difference between the detailed predictions of a multi-zone
photo-chemical code (such as those of Paper I) and the ones presented here are smaller than the effects 
produced by either the merging-induced starburts or by the assumption of an instantaneous initial infall.
The J-K colour is not an useful tool to discriminate among the models.
In general, small (i.e. $M_{acc}\le 0.1 M_{lum}$) or high redshift (i.e. $t_{acc}\le 1$ Gyr) perturbations
do not significantly alter the colours, thus predicting values well within the observed scatter.
Cases \emph{c} and \emph{h} are ruled out, because they predict U-V and V-K colours which are 0.3 mag and 0.5 mag
less than the typical value at a given $M_V$, respectively.
As expected from the analysis made in Paper I, models without infall (i.e. closed box) predict
very blue galaxies. The agreement with the colours is even worse when very recent and massive mergers are allowed 
(e.g. case \emph{m}).
Only one case of massive merger seems to be consistent with observations in the whole mass range, namely the one in which 
$M_{acc}=M_{lum}$ and $t_{acc}=1$ Gyr (models \emph{b, g}). In these particular models, in fact, the merging-induced starburst
increases the mean stellar metallicity (and thus makes the colours redder) and occurs at a sufficiently high redshift 
that it can be considered a continuation of the main star formation episode; in other words, it is like to have
a single burst converting $\sim 1.5 - 2\times M_{lum}$ of gas into stars and evolving passively since then.

On the other hand, when analysing the trend of $[<Mg/Fe>_V]$ versus $M_V$, the presence of a radial
gradient in the abundance ratio should be considered. In fact, we are showing predictions which refer 
to the whole galaxy, whereas the observations are taken within 0.1 effective radius. Moreover,
according to Paper I's best model (see their sec. 3.2), a variation of the order of $\Delta [<Mg/Fe>_V]\sim 0.2 - 0.3$ dex
is expected to occur within one effective radius, with
this ratio increasing outwards.  These predictions have been
recently confirmed by the observed gradients in the galaxies NGC4697 (Mendez
et al. 2005) and NGC821 (Proctor et al., 2005). However, these results are at variance
with the mean value of the gradient slope ($\Delta [<Mg/Fe>_V] \sim 0$) observed by Mehlert et al. (2003)
for a large sample of early-type galaxies in Coma. This topic is
still debated in the literature, given its importance in the study
of the galaxy formation process. At this stage further (either theoretical or observational) efforts are needed,
which go beyond the scope of the present work. 
In any case, in this paper we assume the results of Paper I on
the radial variation of the $[<Mg/Fe>_V]$ ratio. \rm 
When this happens, both models \emph{a} and \emph{f}, their predictions in the $[<Mg/Fe>_V]$-$M_V$ plane move down
in the area where the observations lie (see arrows in Fig. \ref{fig-cmr-d}).
Under the hypothesis that the gradient is unaffected
by merging episodes, it is clear from Fig. \ref{fig-cmr-d} that models with small perturbations
are still in agreement with the data.
In the case of the closed box models, we can see that the predictions can be reconciled
with observations only if a very strong (i.e. $\Delta [<Mg/Fe>_V] > 0.4$) radial gradient in $[<Mg/Fe>_V]$ occurs. However,
this is not predicted for the CB cases and is not confirmed by observations,
which show $\Delta [<Mg/Fe>_V] \le 0.2$ (Mehlert et al., 2003, Mendez
et al., 2005). Among this set of models, only cases \emph{m} and \emph{r}
do not require gradients stronger than those predicted by Paper I's best model. However, they are ruled out
by the photometric analysis.
In more realistic cases the models featuring a merging episode
would face another problem that might make their disagreement with the observations even worse.
In particular, we expect that the newly accreted gas undergoes stronger cooling in the central region,
thus leading to longer star formation in the galactic core with respect to the outskirts.
In this case the radial gradient of $[<Mg/Fe>_V]$
will be even larger than the one predicted
by the best model of Paper I and also at odds with observational results. 
Therefore, the predictions for the central zone of \emph{perturbed} models lie well below the 
observed region, at a given magnitude.

Finally, we remark that all models with a non-negligible merger, exhibit very high metallicity ($[<Fe/H>]\sim 1$, $[<Mg/H>]\sim 0.4$),
when the second burst occurs very late, at variance with observations (e.g. Kobayashi \& Arimoto, 1999,
who give $[<Fe/H>]\sim -0.3$).
Because of the different ages of the two groups of stellar populations 
in the model galaxies (i.e. the old one, made during the first episode of SF, and the young one, made several Gyrs
later during the second burst), we do not consider safe to obtain estimates of the metallicity indices by simply
using a calibration relation, as done in Paper I. Therefore, we avoided this part of the analysis. 
Tantalo \& Chiosi (2004, see also Maraston \& Thomas, 2000) reached conclusions similar to ours
by analysing the spread in the $H_{\beta}-[MgFe]$ plane and the colour evolution. In particular, they found
that the presence of young populations superimposed to the old bulk of stars would be detectable,
at odds with observations. On the other hand, small merging events producing differences in the $\alpha$-enhancement
can explain the observational scatter.
However, in a very recent paper, Thomas et al. (2004) found that
a non-negligible fraction of low-mass ellipticals in their sample
shows evidences of a young (i.e. age $< 2$ Gyr) and metal-rich ($[Z/H]\sim0.2$) stellar population
which represents $\sim 10-20 \%$ of their total stellar mass. These values are independent
from the environment and the typical [$\alpha$/Fe] of this population are
$\sim$ 0.2 dex lower than the bulk of the stars, as expected if the young component
formed out of gas strongly polluted by SNIa (e.g. models \emph{c} and \emph{d}). 

\subsection{Other star formation histories?}

We complete the analysis of both Cases 1 and 2 by considering models
in which the SFH is modified by changing the second burst duration and its star formation efficiency
(see Table 1).
For example, in case \emph{c+$t_{burst}$=0.5} (i.e. model c with $t_{burst}$=0.5 Gyr), 
the colours are still bluer than the fiducial case and very similar to those predicted by model 
\emph{c}, although the predictions regarding the chemical evolution are significantly better. 
In order to improve the agreement between the predictions of the fiducial case and the
observations, we need simultaneously $t_{burst}$=0.1 Gyr, and either a low amount of mass accreted
or a low $t_{acc}$ (i.e. high redshift). In these cases, in fact, the time interval during which new stars
are allowed to form, is short enough to ensure negligible perturbations
in the final abundance estimates.
This result holds even in very extreme cases, namely we
run a model in which $M_{acc}=0.1M_{lum}$, $t_{burst}$=0.1 Gyr and $\nu=1000\rm Gyr^{-1}$, for mergers
occurring at $t_{acc}<4$ Gyr (i.e. z$\sim$1). Despite the large value adopted for $\nu$, in fact,
the SFR is only $\sim 100 M_{\odot} \rm yr^{-1}$, owing to the small amount of gas present
in the ISM. This finding is particularly important because, allowing for mass accretion onto the galaxies, but 
limiting the star formation process to a very short
period, the typical star formation rate 
is of the order of $\sim 40-80 M_{\odot} \rm yr^{-1}$ ($\sim 4-6 M_{\odot} \rm yr^{-1}$) 
for model \emph{b+$t_{burst}$=0.1} (\emph{d+$t_{burst}$=0.1}). Therefore, these models 
cannot convert into stars all the gas they accreted, and might end up with a cold and massive ISM
(even $\sim 0.4 - 0.5 M_{lum}$ in cases with $M_{acc}=M_{lum}$), which is not observed in present-day ellipticals. 
Allowing for a very high SF efficiency it is helpful in removing all the cold gas and, at the same time, 
it shows the robustness
of our conclusions for a large range of SFR.
\rm

Finally, we consider cases in which
also the SFH of the initial burst deviates significantly from the
fiducial model, in order to complete the study of the parameter
space for Cases 2 and 3.
We concentrate on cases in which the galaxy accretes a gas mass  
$M_{acc}=M_{lum}$, since in the previous sections
we have shown that models with $M_{acc}=0.1\,M_{lum}$
can be considered as small perturbations (see the ratio $R_*$ in Table 1).
For what concerns Case 2, the aim is similar to what done in Sec. 3.2., namely
we try to have a new best-fitting model for the [Mg/Fe] vs.
mass relation. 
In particular, we explore the cases in which
we keep $\tau$ fixed and change $\nu$.
Values of $\nu = 35-45\rm\, Gyr^{-1}$ during
both the main SF episode and the starburst of model g improve
only mildly the agreement with observations.
A significant increase in the predicted value
for $[<Mg/Fe>_V]$ (i.e. 0.363 dex) is attained
when $\nu = 55\rm\, Gyr^{-1}$ (model \emph{u}). In this case, 
we predict the ratio $R_*$ to be $\sim 0.5$, therefore
it can be regarded as a model similar to \emph{g}, but with
a more intense SFH during the second burst.
\rm
The problem is that,
however, the colours start to deviate from
the observed relations (U-V=1.67, V-K=3.48 for
$M_V = -22.2$ mag) and the Fe abundance
becomes quite high ($[<Fe/H>_V]=0.700$).

We also tried another exercise, namely
we abandoned our self-consistent evaluation of the time at which the
galactic wind occurs, based on a detailed treatment
of the SN remnant evolution, and fixed it a priori.
Here we present the results for model g, with
$t_{gw}=0.25$ Gyr.
In this case, although the predicted abundance ratios
(namely $[<Mg/Fe>_V]=0.433$, $[<Fe/H>_V]=0.414$)
are fairly consistent with observations, the colours become too blue.
In fact, we predict U-V=1.22 and V-K=2.83 for a total V magnitude $M_V = -21.0$ mag.

This ad hoc model can be seen as a possible way to extend model \emph{g}
to Case 2. In fact, 
owing to the adopted self-consistent treatment for the development of the galactic
wind, model \emph{g} cannot reach $R_* \sim 1$. Here, instead, 
we reduce the SF timescale during the initial burst (i.e. $t_{gw}$) by a factor of $\sim 4$;
therefore we have $R_* \sim 1$.
\rm

\begin{figure}
\epsfig{file=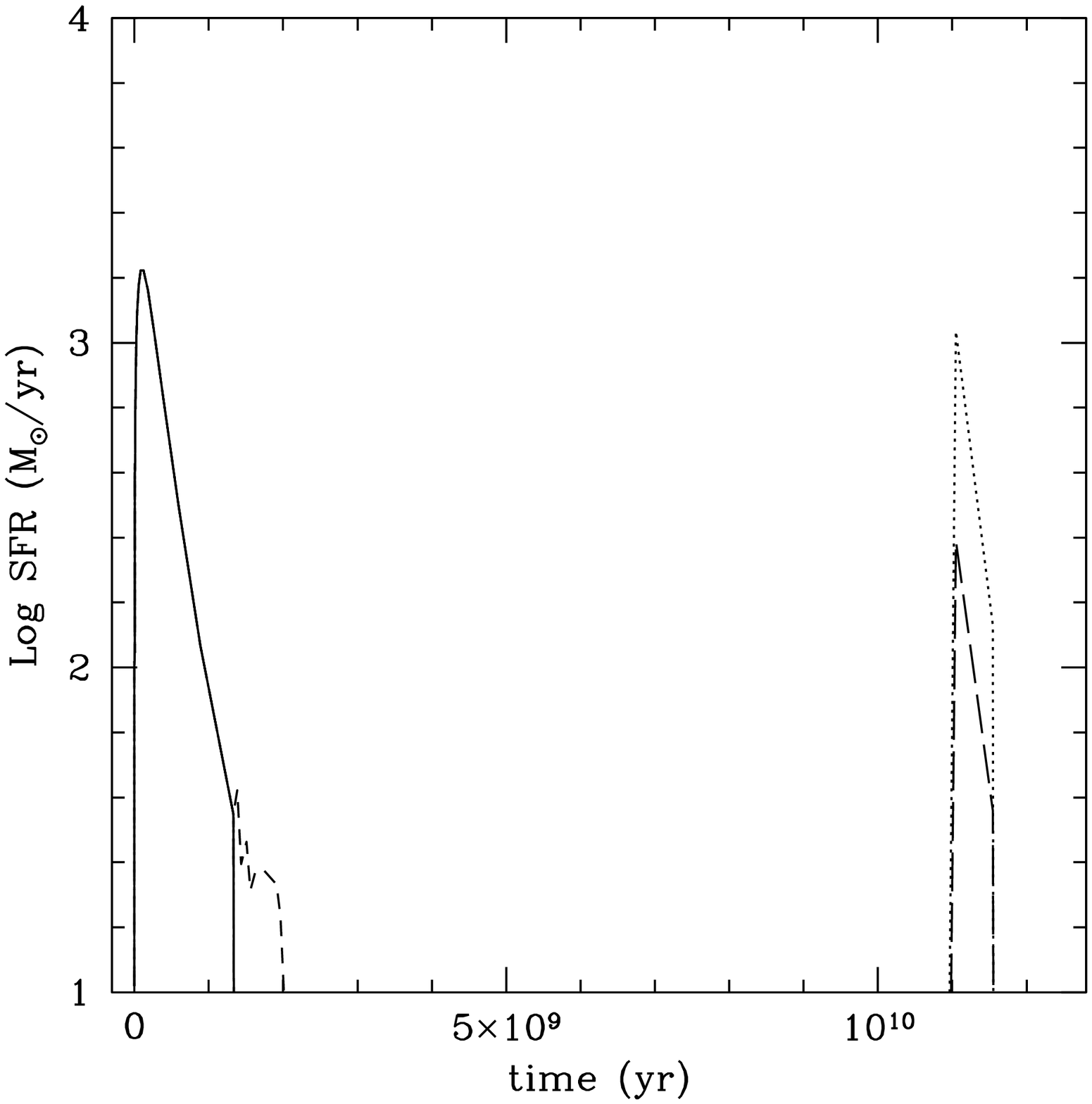, height=8cm, width=8cm}
\epsfig{file=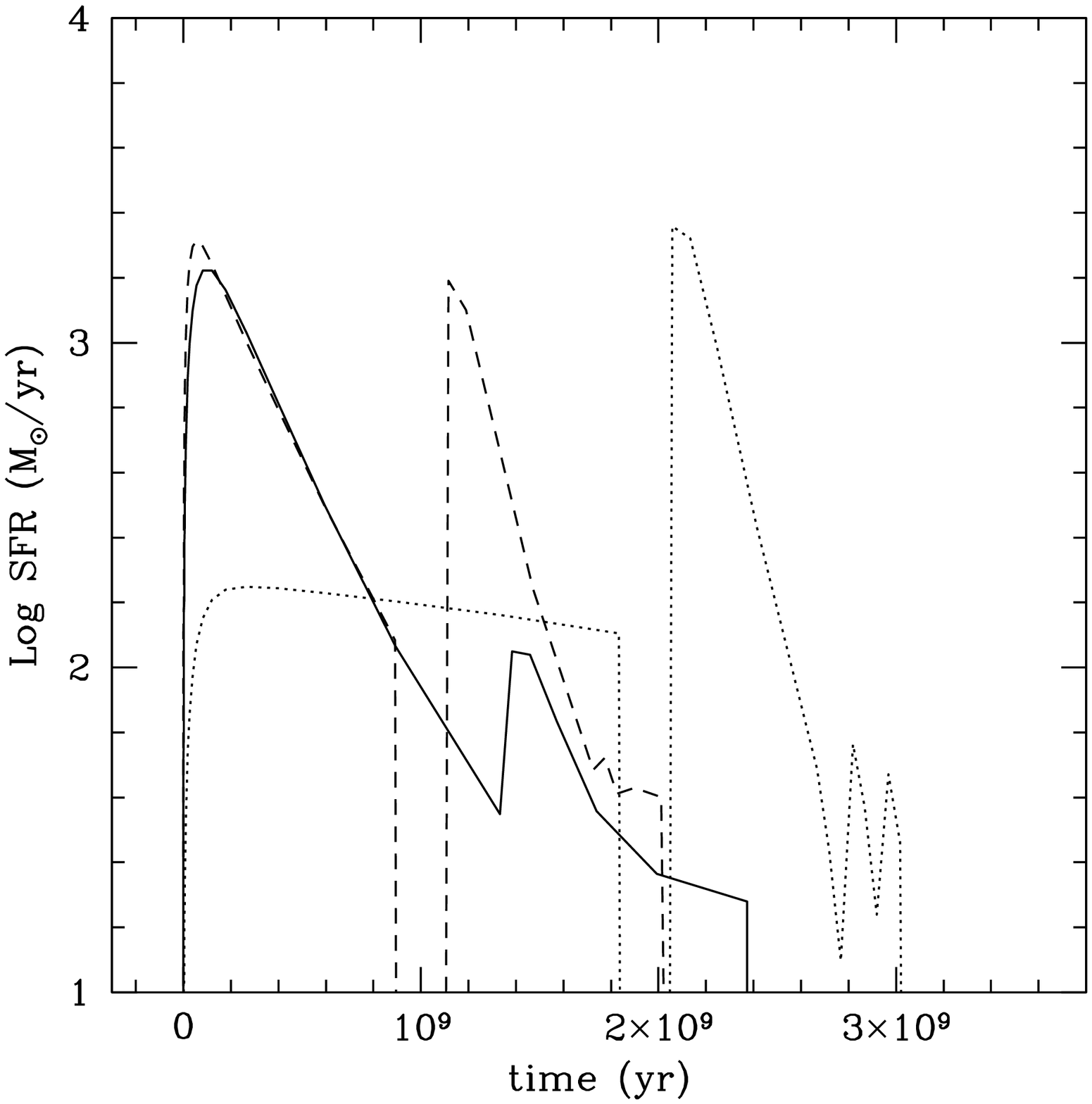, height=8cm, width=8cm}
\caption{Star formation rate for some representative cases.
\emph{Upper panel:}
Solid: Paper I's best model (\emph{f}). Dotted: The SFH during the main burst and during the merger
have similar intensity and duration (as in Case 2, model \emph{h}).
Two models in which $M_{acc}=0.1M_{lum}$ are shown by dashed (model \emph{i}) and
long-dashed (model \emph{j}) lines, respectively.
All these models have the same SFH during the initial burst.
\emph{Lower panel:}
Solid: The SFH during the main burst is more intense than during the merger
(model \emph{g}). Dashed: The SFH during the main burst and during the merger
have similar intensity and duration (as in Case 2, model \emph{u}).
Dotted: The SFH during the main burst is less intense than during the merger
(as in Case 3, model \emph{v}). 
}
\label{fig-sfr}
\end{figure}

Concerning Case 3, we start by increasing the SF efficiency during
the second burst. For instance see the model \emph{c+$\nu$=20} 
(i.e. model \emph{c} with $\nu =20 Gyr^{-1}$,  which leads to $R_* \sim 2$\rm ), which improves the agreement between
the predictions regarding the chemical properties, although
the colours are still bluer than the fiducial case.
Finally, we present model \emph{v}, an illustrative case for a $10^{12}M_{\odot}$ galaxy
in which we vary the parameters in order
to have a mild and prolonged SF before the onset
of the merger-induced burst (occurring at 2 Gyr). 
This corresponds to the region of parameter
space in which $\tau$ varies in the interval [0.5, 10] Gyr
and $\nu \le 20 \rm \, Gyr^{-1}$ (during the first burst).
In particular, we chose $\tau=3$ Gyr, $\nu = 2.2 \rm\, Gyr^{-1}$
and $\nu = 110 \rm\, Gyr^{-1}$ during the burst.
A wind developes only after the starburst.The model
exhibits $[<Mg/Fe>_V]=0.416$, U-V=1.48 and V-K=3.25 (for a $M_V = -21.8$ mag),
values which are pretty close to Paper I's best model results. Nevertheless,
the SF produces a very high Fe enrichment in the 
bulk of the stars ($[<Fe/H>_V]=0.839$) which is not
observed. In the hypothesis that this kind
of galactic model can be fairly represented by a SSP
(but see Pipino, Matteucci \& Chiappini 2005, in preparation),
this would translate into the \emph{mean iron} Lick index $<Fe>=4.5$
(once Worthey's 1994 simple stellar populations are adopted, see Paper I).

The effect of moving the epoch of the mergers to $t_{acc}\ge$ 5 Gyr
is to worsen the results, since the colours become progressively bluer
(e.g. V - K =3.13 for $M_V = -22.1$ mag).

Given its $R_* \sim 2$ and the behaviour of its SFH, \rm
to some extent, this is a complementary result of what Matteucci \& Pipino (2005) have shown for PopIII stars
in ellipticals (see also Gibson 1996 for a more extended
discussion on the bimodal SF in ellipticals) by changing the IMF
at early epochs.  
In any case, even a short pre-enrichment of the bulk of the stellar populations,
produce either too red galaxies or abundance patterns
in disagreement with observations.

In Fig. \ref{fig-sfr} we compare the star formation 
histories of some models, chosen as the most representative cases shown in the previous
sections.
In the upper panel we compare the Paper I best model (\emph{f}) SFH  with
those from the cases \emph{h, i, j}.
In the lower panel, the solid line is taken from model \emph{g}: the main burst clearly
dominates over the merger-induced SF episode. Case 2 is represented
by model \emph{u} (dashed line): the two SF episodes have a very close
intensity and shape. They are separated by a short episode of galactic wind.
Finally, a model in which the SF during the merger-induced burst exceeds
the initial episode (Case 3, model \emph{v}) is shown by a dotted line.
\rm

\subsection{Other parameters effect}

At this point it is useful to discuss other possible sources of degeneracy in our model results.
First of all, it should be noticed that all the cases but the two Paper I's best models (labelled as \emph{Paper I} in Table 1),
fail in reproducing the present-day SNIa rate observed in ellipticals ($0.18\pm 0.06$ SNu, Cappellaro et al 
1999). For example, model \emph{c} predicts an explosion rate of $\sim$ 1 SNu, whereas
model \emph{b} gives 0.48 SNu. The latter value, however, is only marginally consistent
with the rate of $0.3\pm 0.1$ SNu given by Cappellaro et al. (1999) for ellipticals
bluer than the average. 
The reason for this disagreement is that the parameter $A$ in eq. (1), 
is set equal to 0.18 in order to have a good fit in the case of models \emph{a} and \emph{f}. For comparative
purposes, we run a model in which we use a lower value for A (Table 1, model \emph{b+A=0.06}),
namely we chose A=0.06 so that the present-day SNIa rate for model \emph{b} becomes 0.19 SNu.
This leads to luminosities and colours essentially unchanged, whereas the $[<Mg/Fe>_V]$ ratio rises
toward values more similar to Paper I's best model predictions,
as a consequence of the lower Fe abundance in the ISM when the merger-induced starburst occurs.
However, a systematic variation of A with the properties
of the merger, would be in contrast with the \emph{a-priori} definition of A. 

Another possible solution can be a variation either with time or metallicity of the fraction 
of binary systems (e.g. De Donder \& Vanbeveren, 2002). It should be remarked, however, that,
in elliptical galaxies, the metallicity increases very fast in the first hundreds of Myr, reaching
the solar value at the time of the galactic wind, and being higher than solar in the wind regime. Therefore,
the \emph{final} (i.e. present-day) value for A should be achieved well before 
the occurrence of the second burst of star formation.
In any case, we do not consider this as a viable solution for reconciling star formation histories
ruled by random gas accretion episodes with the observed features of elliptical galaxies.

We also tried to change the IMF, by using a typical flat one (Arimoto \& Yoshii 1987, models \emph{a+AY87} and \emph{b+AY87}). 
This IMF was ruled out in Paper I because it predicts a too high metallicity in the stellar populations
and too high values for $[<Mg/Fe>_V]$. The merging-induced starburst, when occurring in these models,
pushes the $[<Mg/Fe>_V]$ down to a reasonable value for galaxies of that given mass. On the other hand,
the Fe and Mg abundances exceed the solar value by a factor of ten, at variance with what is suggested by 
the observations.
The possibility for a flat IMF during a merging process was explored by Thomas (1999),
who concluded that a significantly flat exponent (i.e. $x< 0.8$) is needed to satisfy
the constraints on $[<Mg/Fe>_V]$.

Finally, we tested the effect of changing the composition of the accreted gas. In particular we focused
on a solar composition, as given by Anders \& Grevesse (1989). 
When this option is applied to model \emph{d}, the differences with the predictions
shown in Table 1 are negligible. This is a straightforward consequence
of the fact that the total mass of the accreted gas which is turned into stars is very small compared to the mass of
the dominant stellar population. A stronger effect is expected for models in which
$M_{acc}=M_{lum}$. In this case, in fact, model \emph{c} would exhibit $[<Mg/Fe>_V]=0.258$, owing to 
the great importance of the youngest stellar populations in the weighted average.
Also the colours become redder (by $\sim$ 0.1 mag), but not enough to match the CMRs. On the other hand, the effect 
of assuming $Z_{acc}=Z_{\odot}$ on model \emph{b} predictions is milder for the chemical
aspect, whereas the colours are $\sim$ 0.1 mag redder.
This implies that, in any case, the time of the occurrence of the merger-induced starburst is more important than the chemical composition 
of the accreted gas.

\section{Conclusions}

In this paper we completed the analysis of the photochemical 
properties of elliptical galaxies begun in Paper I, where we tested the high redshift
formation of ellipticals. 
We showed that a photo-chemical diagnostic, namely 
the CMRs and the $[<Mg/Fe>_V]$-$M_V$, is needed in order to verify whether star formation histories 
differing from one of the simple \emph{monolithic} one lead
to the same good agreement with observations. This diagnostic allowed us to
quantify Paper I's best model response to perturbations, represented by merger-induced starbursts.
Here we summarise our main conclusions:
\begin{itemize}

\item Paper I's best model seems to preserve the consistency with observations, 
within the observed scatter, only when small merging events occur.
These results reinforce Paper I findings, namely that the star formation process must be
more efficient and faster in the more massive galaxies in order to reproduce the largest possible
set of optical observables. The measured scatter in these relations allows 
only small and early \emph{perturbations}, i.e. $M_{acc}\sim 0.1M_{lum}$ and $z_{acc}\ge$3.

\item Either bursts lasting more than 0.5 Gyr and involving
a gas mass comparable to the mass already transformed into stars during the first episode of star formation,
or occurring at low redshift (i.e. z=0.2) are ruled out. Only cases in which
the burst duration is limited to 0.1 Gyr are marginally consistent with observations, 
if $t_{acc}$ is small.

\item Furthermore, these merging models fail in matching the above relations in the case in which 
the initial infall hypothesis is relaxed, and the galaxies form through 
the classical monolithic paradigm (e.g. Larson 1974). Among these models,
galaxies accreting a small amount of gas at high redshift produce a spread in the model results,
with respect to Paper I's best model, which is consistent with the observational scatter
of the CMRs, but they fail in reproducing the $[<Mg/Fe>_V]$-$\sigma$ relation.

\item \rm On the other hand, models in which the SFH during the merger leads to
the creation of many more stars with respect to what happened during the \emph{initial}
burst, do not match the observational constraints on ellipticals. \rm

\item The possibility of intermediate cases is appealing, but, because of the tightness of the observed 
colour-magnitude, mass-metallicity \emph{and} $[<Mg/Fe>_V]$-$\sigma$ relations,
it seems to be inconsistent with a systematic presence of merging episodes, given the stochastic
nature of the accretions along the galactic lifetime.
\end{itemize}
Finally, we remark that the strongest constraints to the galactic
formation mechanism and to the occurrence of late time merging events 
are provided by the chemical abundances.

\section*{Acknowledgments} 

Useful discussions with F. Calura, C. Chiappini, J. Koeppen, R. Maiolino and S. Samurovic are acknowledged. 
We thank J. Danziger for a careful reading of the paper.
The work was supported by MIUR under COFIN03 prot. 2003028039.
We thank the referee, whose comments improved the quality of the paper.


\begin{thebibliography}{}

\small
\bibitem []{}Anders, E., $\&$ Grevesse, N., 1989, Geochim. Cosmochim. Acta, 53, 197
\bibitem []{}Arimoto, N., $\&$ Yoshii, Y. 1987, A$\&$A, 173, 23 
\bibitem []{}Bernardi, M., Sheth, R.K., Annis, J., et al. 2003, AJ, 125, 1882
\bibitem []{}Beuing, J., Bender, R., Mendes de Oliveira, C., Thomas, D., Maraston, C., 2002, A$\&$A, 395, 431
\bibitem []{}Bower, R.G., Lucey, J.R., Ellis, R.S., 1992a, MNRAS, 254, 589
\bibitem []{}Bower, R.G., Lucey, J.R., Ellis, R.S., 1992b, MNRAS, 254, 601
\bibitem []{}Calura, F., Matteucci, F., 2004, MNRAS, 350, 351
\bibitem []{}Calura, F., Matteucci, F., Menci, N., 2004, 353, 500
\bibitem []{}Cappellaro, E., Evans, R., Turatto, M., A$\&$A, 1999, 351, 459
\bibitem []{}Carollo, C.M., Danziger, I.J., $\&$ Buson, L. 1993, MNRAS, 265, 553
\bibitem []{}Cimatti, A., Daddi, E., Renzini, A., Cassata, P., Vanzella, E., Pozzetti, L., 
Cristiani, S., Fontana, A., Rodighiero, G., Mignoli, M., Zamorani, G. 2004, Nature, 430, 184
\bibitem []{}Davies, R.L., Sadler, E.M., $\&$ Peletier, R.F., 1993, MNRAS, 262, 650
\bibitem []{}de Donder, E., Vanbeveren, D. 2002, NewA, 7, 55
\bibitem []{}Ellis, R.S., Smail, I., Dressler, A., Couch, W.J., Oemler, A.Jr., Butcher, H., $\&$ Sharples,
R.M., 1997, ApJ, 483, 582
\bibitem []{}Faber, S.M., Worthey, G., $\&$ Gonzalez, J.J. 1992, in IAU Symp. n.149,
eds. B. Barbuy $\&$ A. Renzini, p. 255
\bibitem []{}Gibson, B.K., 1996, MNRAS, 278, 829 
\bibitem []{}Gibson, B.K., 1997, MNRAS, 290, 471
\bibitem []{}Gonzalez, J.J., 1993, PhD thesis, Univ. of California
\bibitem []{}Granato, G.L., De Zotti, G., Silva, L., Bressan, A., Danese, L. 2004, ApJ, 600, 580
\bibitem []{}Greggio, L., $\&$ Renzini, A. 1983, A$\&$A, 118, 217
\bibitem []{}Heckman, T.M. 2002, Extragalactic Gas at Low Redshift, ASP Conference Proceedings Vol. 254. Eds. J.S. Mulchaey \& J. Stocke.  
San Francisco: Astronomical Society of the Pacific, p.292
\bibitem []{}Jimenez, R., Padoan, P., Matteucci, F., Heavens, A.F., 1998, MNRAS, 299, 123
\bibitem []{}Kauffmann, G., $\&$ Charlot, S. 1998, MNRAS, 294, 705
\bibitem []{}Kaviraj, S., Devriendt, J.E.G., Ferreras, I., Yi, S.K., 2005, MNRAS, 360, 60
\bibitem []{}Kobayashi, C., $\&$ Arimoto, N. 1999, ApJ, 527, 573
\bibitem []{}Kuntschner, H. 2000, MNRAS, 315, 184
\bibitem []{}Kuntschner, H., Lucey, J.R., Smith, R.J., Hudson, M.J., Davies, R.L.  2001, MNRAS, 323, 615
\bibitem []{}Larson, R.B., 1974, MNRAS, 166, 585
\bibitem []{}Maraston, C., Thomas, D., 2000, ApJ, 541, 126
\bibitem []{}Matteucci, F. 1994, A$\&$A, 288, 57
\bibitem []{}Matteucci, F., $\&$ Greggio, L., 1986, A$\&$A, 154, 279
\bibitem []{}Matteucci, F., \& Pipino, A., 2005, MNRAS, 357, 489
\bibitem []{}Matteucci, F., Ponzone, R., Gibson, B.K., 1998, A$\&$A, 335, 855
\item []{}Matteucci, F., $\&$ Tornambe', A., 1987, A$\&$A, 185, 51
\bibitem []{}Mehlert, D., Saglia, R.P., Bender, R., $\&$ Wegner, G. 2000, A$\&$AS, 141, 449
\bibitem []{}Mehlert, D., Thomas, D., Saglia, R.P., Bender, R., $\&$ Wegner, G. 2003, A\&A, 407, 423
\bibitem []{}M{\' e}ndez R.~H., Thomas D., Saglia 
R.~P., Maraston C., Kudritzki R.~P., Bender R., 2005, ApJ, 627, 767
\bibitem []{}Meza, A., Navarro, J.F., Steinmetz, M., Eke, V.R. 2003, ApJ, 590, 619
\bibitem []{}Nomoto, K., Hashimoto, M., Tsujimoto, T., Thielemann, F.K., Kishimoto, 
N., Kubo, Y., Nakasato, N., 1997, Nuclear Physics A, A621, 467
\bibitem []{}Peebles, P.J.E. 2002, in A New Era in Cosmology, ASP
Conference Series, S. Francisco, eds. N.Metcalfe $\&$ T.Shanks, p.351 
\bibitem []{}Pipino, A., Matteucci, F., 2004, MNRAS, 347, 968 (Paper I)
\bibitem []{}Pipino, A., Kawata, D., Gibson, B.K., Matteucci, F., 2005, A\&A, 434, 553
\bibitem []{}Proctor et al. , 2005, astro-ph/0506523
\bibitem []{}Rusin, D., Kochanek, C.S., Falco, E.E., Keeton, C.R., McLeod, B.A., Impey, C.D., Lehar, J.,
Munoz, J.A., Peng, C.Y., Rix, H.W. 2003, ApJ, 587, 143
\bibitem []{}Saglia, R.P., Maraston, C., Greggio, L., Bender, R., $\&$ Ziegler, B. 2000, 360, 911
\bibitem []{}Salpeter, E.E., 1955, ApJ, 121, 161
\bibitem []{}Shioya, Y., Bekki, K., 1998, ApJ, 504, 42
\bibitem []{}Stanford., S.A., Eisenhardt, P.R., $\&$ Dickinson, M. 1998, ApJ, 492, 46
\bibitem []{}Steinmetz, M., Navarro, J.F. 2002, NewA, 7, 155
\bibitem []{}Tantalo, R., Chiosi, C., 2004, MNRAS, 353, 405
\bibitem []{}Thielemann, F.K., Nomoto, K., Hashimoto, M. 1996, ApJ, 460, 408 
\bibitem []{}Thomas, D., 1999, MNRAS, 306, 655
\bibitem []{}Thomas, D., Greggio, L., $\&$ Bender, R., 1999, MNRAS, 302, 537
\bibitem []{}Thomas, D., Maraston, C., $\&$ Bender, R., 2002, Ap$\&$SS, 281, 371
\bibitem []{}Thomas, D., Maraston, C., $\&$ Bender, R., 2003, MNRAS, 339, 897 
\bibitem []{}Thomas, D., Maraston, C., $\&$ Bender, R., Mendes de Oliveira, C., 2005, ApJ, 621, 673
\bibitem []{}Trager, S.C., Faber, S.M., Worthey, G., Gonzalez, J.J., 2000a, AJ, 119, 1654
\bibitem []{}Trager, S.C., Faber, S.M., Worthey, G., Gonzalez, J.J., 2000b, AJ, 120, 165
\bibitem []{}Trager, S.C., Worthey, G., Faber, S.M., Burstein, D., Gonzalez, J.J., 1998 ApJS, 116, 1
\bibitem []{}van de Ven, G., van Dokkum, P.G., Franx, M. 2003, MNRAS, 344, 924
\bibitem []{}van den Hoek, L.B., Groenewegen, M.A.T. 1997, A$\&$AS, 123, 305
\bibitem []{}van Dokkum, P.G., \& Ellis, R.S., 2003, Apj, 592L, 53
\bibitem []{}van Dokkum, P.G., $\&$ Franx, M. 1996, MNRAS, 281, 985
\bibitem []{}van Dokkum, P.G., Franx, M., Kelson., D.D., Illingworth, G.D., Fisher, D., Fabricant, D. 1998, ApJ, 500, 714
\bibitem []{}Weiss, A., Peletier, R.F., Matteucci, F. 1995, A$\&$A, 296, 73
\bibitem []{}Whelan, J., Iben, I. Jr. 1973, ApJ, 186, 1007
\bibitem []{}White, S.D.M., $\&$ Rees, M.J., 1978, MNRAS, 183, 341
\bibitem []{}Worthey, G. 1994, ApJS, 95, 107
\bibitem []{}Worthey, G., Faber, S.M., $\&$ Gonzalez, J.J. 1992, ApJ, 398, 69


\end{thebibliography}
\end{document}